\documentclass[aps,prl,twocolumn,showpacs,floatfix]{revtex4-1}
\usepackage{amsmath,amsfonts,bm,color,amssymb}
\usepackage{graphicx,epsfig,overpic,graphics,epstopdf}

\usepackage{epstopdf}

\begin{document}
\title{Experimental Realization of an Incompressible Newtonian Fluid in Two Dimensions}

\date{\today}
\author{Zhiyuan Qi}
\author{Cheol Soo Park}
\author{Matthew A.~Glaser}
\author{Joseph E.~Maclennan}
\author{Noel A.~Clark}
\affiliation{Department of Physics and the Liquid Crystal Materials Research Center, University of Colorado, Boulder, Colorado, 80309, USA}
\date{\today}

\begin{abstract}
The Brownian diffusion of micron-scale inclusions in freely suspended smectic~A liquid crystal films a few nanometers thick and several millimeters in diameter depends strongly on the air surrounding the film. Near atmospheric pressure, the three-dimensionally coupled film/gas system is well described by Hughes/Pailthorpe/White hydrodynamic theory but at lower pressure ($p \alt 70$ torr), the diffusion coefficient increases substantially, tending in  high vacuum toward the two-dimensional limit where it is determined by film size. In the absence of air, the films are found to be a nearly ideal physical realization of a two-dimensional, incompressible Newtonian fluid.
\end{abstract}

\maketitle
Theoretical hydrodynamics has progressed through the invention of a series of abstract fluids (perfect, inviscid, incompressible, and so on) that enable the tractable description of certain physical aspects of three-dimensional ($3$D) fluid systems \cite{Lamb1916}. Among the most useful of these idealizations has been that of the incompressible Newtonian fluid, which models the low-Reynolds number flow of simple and weakly-associated liquids, for example. While there are many physical realizations of such fluids in $3$D, there have been none which satisfy the basic requirements in $2$D, i.e., which is homogeneous in density and viscosity, and obeys the laws of conservation of mass, energy, and momentum. Currently studied $2$D fluids include soap films \cite{Goldburg2002}, which are highly compressible in-plane due their facile response to stress (resulting in thickness changes); and few-nanometer thick, freely suspended, fluid smectic liquid crystal films \cite{Young1978} which, by virtue of their lamellar structure, are quantized in thickness to an integral number of layers, stabilizing hydrodynamic parameters such as density and viscosity to an extent comparable to that of $3$D fluids. Both systems exchange momentum and energy with a surrounding gas but the low vapor pressure~\cite{Deschamps2008,Poole2014} of smectic films enables the possibility of pressure reduction to the microtorr regime and thereby the approach to, and study of, the ideal incompressible , isotropic, Newtonian limit of 2D fluids (2DIIN limit). The experiments on smectic films reported here explore the evolution to this hydrodynamic regime as the surrounding gas pressure is reduced and investigate the anomalies arising from  reduced dimensionality in this limit.
  
Hydrodynamic behavior in $2$D has received extensive theoretical attention \cite{Goldenfeld2007} and is of broad interest in the context of $2$D flows in $3$D systems, ranging from wires falling in $3$D viscous fluids \cite{White1946} to the large scale motion of oceans and the atmosphere~\cite{Boffetta2012}. Also there is increasing interest in the flow of $2$D films \emph{per se} in connection with understanding of the dynamical behavior of defects \cite{Muzny1992,Pargellis1992}, textures \cite{Lee2006,Dolganov2014} and inclusions \cite{Nguyen2010,Schulz2014,Qi2014}, and of transport in biological membranes \cite{Simons1997,Hormel2014}, all of which benefit from experimental information at or near the $2$DIIN limit. As an example, the recent experiments of May et al. \cite{May2012} reveal a dramatic alteration of the shape dynamics of free-floating bubbles as a result of a partial suppression of in-plane compressibility.
  
The coupling of an incompressible Newtonian $2$D fluid to the surrounding media was first treated by Saffman and Delbr\"uck (SD) \cite{Saffman1975}. They developed a continuum hydrodynamic model to describe the mobility $\mu$ of an inclusion of radius $a$, in a fluid film of viscosity $\eta$, surrounded by bulk fluid of different viscosity, $\eta^\prime$. They showed that flow of the film about a moving inclusion is limited to a radius on the film of characteristic dimension $l_S = \eta^\prime h/\eta$, the Saffman length. SD treated the case $a < l_S$, finding that $\mu$ is controlled by the film viscosity and the film exhibits $2$D flow as if bounded at $l_S$. Hughes, Pailthorpe, and White (HPW), \cite{Hughes1981} extended SD theory to describe inclusions of arbitrary radius, showing that for large inclusions ($a > l_S$) $\mu$ is determined by friction with the surrounding fluid, exhibiting something more like $3$D Stokes behavior ($\mu \sim 1/a$) \cite{Nguyen2010}.  Aspects of these SD/HPW predictions have since been tested in experiments by several groups \cite{Nguyen2010,Cheung1996,Petrov2012}. In the absence of surrounding fluid the film flow behavior should be $2$D, marked by extremely long ranged (logarithmic) hydrodynamic interaction and inclusion mobilities that depend logarithmically on system size.

In this Letter, we describe the Brownian dffusion of silicone oil droplet inclusions in smectic films as the ambient air pressure is varied from $633$ torr down to $10^{-4}$ torr. The experiments confirm that, while at atmospheric pressure the mobilities are limited by the surrounding gas, in the high vacuum $2$DIIN limit the hydrodynamics  are controlled by film size. In addition, predictions in the $2$DIIN limit describe well the dependence of the mobility of inclusions on distance from the film boundary.

Since friction from the air plays such an important role in determining the hydrodynamic behavior of inclusions in smectic films, understanding how the inclusion-air interactions can be tuned by varying the ambient pressure is of fundamental interest. As the air pressure is reduced, the mean free path $\lambda$ of the air molecules is expected to increase, as indicated in Fig.~\ref{fig2}(d). At sufficiently low pressure, the surrounding air can not be regarded simply as an incompressible, continuous fluid and the well-established SD/HPW model based on low Reynolds number hydrodynamics can no longer be used to predict the mobilities of inclusions. Here we explore the effects of varying the ambient air pressure on the Brownian motion of inclusions in FSLC films, showing that as the air is removed, the system evolves from a pseudo-$3$D regime where coupling to the air is dominant to a regime in which the hydrodynamics are determined by confinement at the boundaries, as predicted for an ideal $2$D fluid.

\begin{figure}[tbh]
\centering
\includegraphics[width=7.5cm]{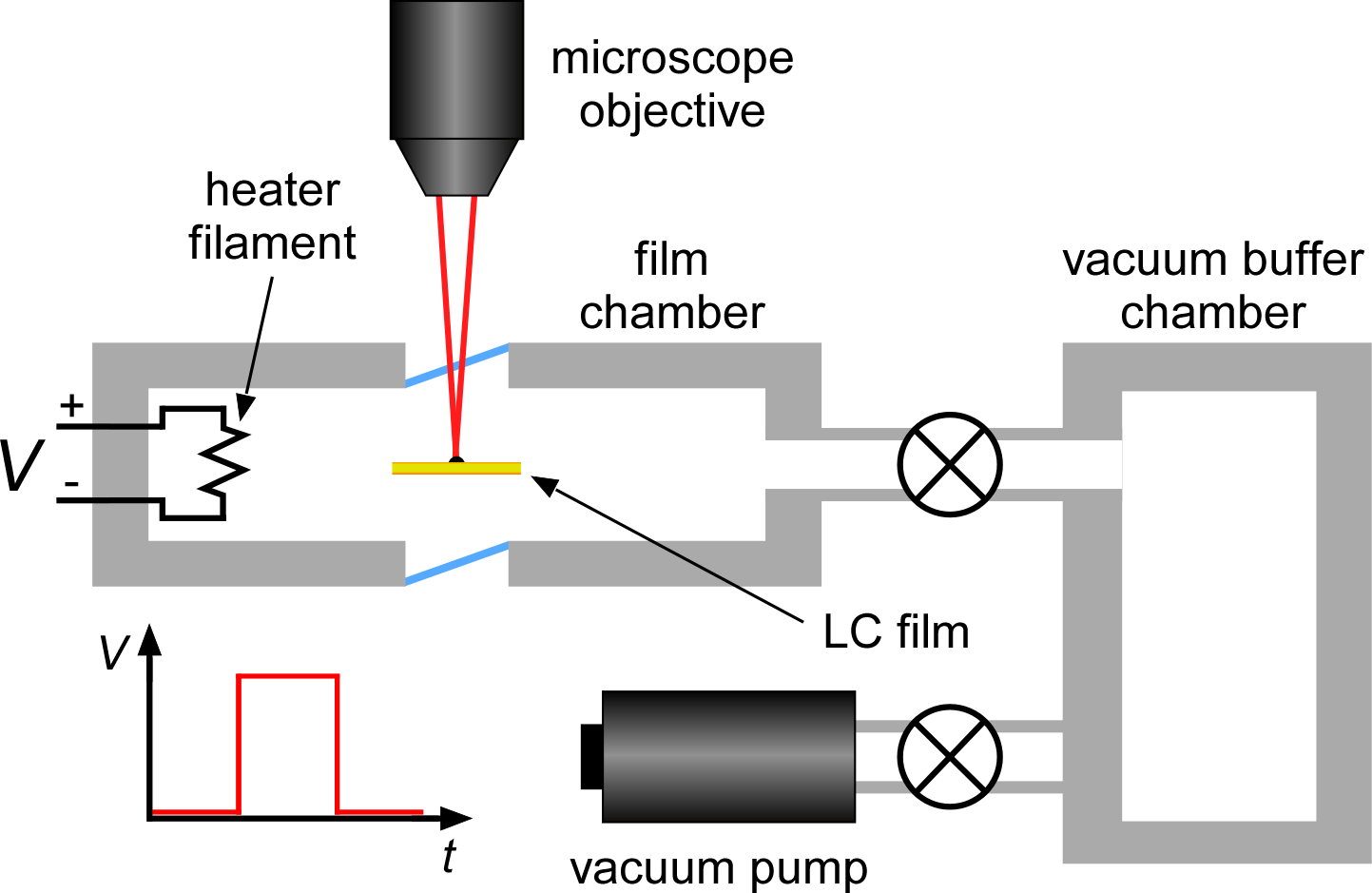}
\caption{\label{fig1} Experimental apparatus for observing inclusions in smectic liquid crystal films at low pressure. A resistive filament coated with silicone oil is briefly heated with an electric current to generate an oil vapor, part of which then condenses as droplets on the film. The buffer chamber shields the film chamber from sudden changes in pressure.}
\end{figure}

Homogeneous FSLC films a few molecular layers thick are robust preparations \cite{Young1978} provide an ideal platform for studying hydrodynamics in reduced dimensions \cite{Muzny1992}. In  previous experiments, we described the Brownian motion of silicone oil droplets embedded in such films with the ambient air at atmospheric pressure \cite{Qi2014}. These droplets form lens-shaped inclusions are insoluble in liquid crystal and whose size remains constant over long time intervals, typically for more than half an hour, which far exceeds the time required to perform a typical measurement.

The liquid crystal material used in our experiment is $8$CB (4\ensuremath{'}-n-octyl-4\ensuremath{'}-cyanobiphenyl), which is in the fluid smectic~A phase at room temperature. The saturation vapor pressure of $8$CB is very low (around $10^{-7}$ torr \cite{Deschamps2008}), and we are able to maintain stable films of constant thickness over a wide range of air pressures (from atmospheric pressure to $10^{-6}$~torr), enabling us to study the microrheology of inclusions in the film over a wide range of Knudsen number ($\lambda/2R$, the reduced mean free path). The density and viscosity of $8$CB are $\rho \approx 0.96 \; \mathrm{g/cm}^3$~\cite{Leadbetter1976} and $\eta = 0.052~{\rm Pa} \cdot {\rm s}$ \cite{Schneider2006} respectively. Each smectic layer is $3.17\,{\rm nm}$ thick \cite{Davidov1979}. Freely suspended films were formed by spreading a small amount of the liquid crystal across a $4\,{\rm mm}$-diameter hole in a glass cover slip and were then observed using reflected light video microscopy. The film thickness $h$, an integral number $N$ of smectic layers (typically $2 \le N \le 6$), is determined precisely by comparing the reflectivity of the film with black glass \cite{Sirota1987}. A resistive filament coated in silicone oil is then electrically heated in order to generate an oil vapor, some of which makes its way to the film where it eventually condenses and forms visible droplets such as those shown in Fig.~\ref{fig2}a, with radii between $2$ and $50\, \mu$m. A double-sealed rotary pump is then used to reduce the pressure to $3 \times 10^{-3}$~torr, after which a turbo pump is used to get the film chamber down to $10^{-4}$ torr. The shape and thickness of the oil droplets is measured by analyzing the interference fringes (see Fig.~\ref{fig2}b) formed in monochromatic light \cite{Schuring2002}.

Once oil droplets appear on the film, the chamber is carefully tilted in order to maneuver a droplet of desired radius $a$ into the center of the film (of radius $R$), after which the film is leveled to minimize gravitational drift, enabling us to capture several thousand images at a video frame rate of $24$~fps while the droplet is in the field of view and far from the film boundaries, as shown schematically in Fig.~\ref{fig2}(c).

After reducing the pressure of the film chamber to $10^{-4}$ torr, closing the valve between the pump and the vacuum buffer chamber allows us to maintain quasi-constant pressure around the film for dozens of minutes, during which we are able to make video recordings of droplet motion. The pressure is then gradually increased by injecting small amounts of air into the system, allowing us to obtain a series of movies showing the Brownian motion of the droplet as the chamber pressure increases in steps from $3\times 10^{-4}$ torr to $633$ torr. These videos are decomposed into sequential images and the size and position of the droplet are determined using algorithms based on Canny's method for edge detection \cite{Canny1986} and Taubin's method for object identification \cite{Taubin1991}. The diffusion coefficient is obtained after analytically removing any drift \cite{Nguyen2010}.

\begin{figure}[tbh]
\centering
\includegraphics[width=7.0cm]{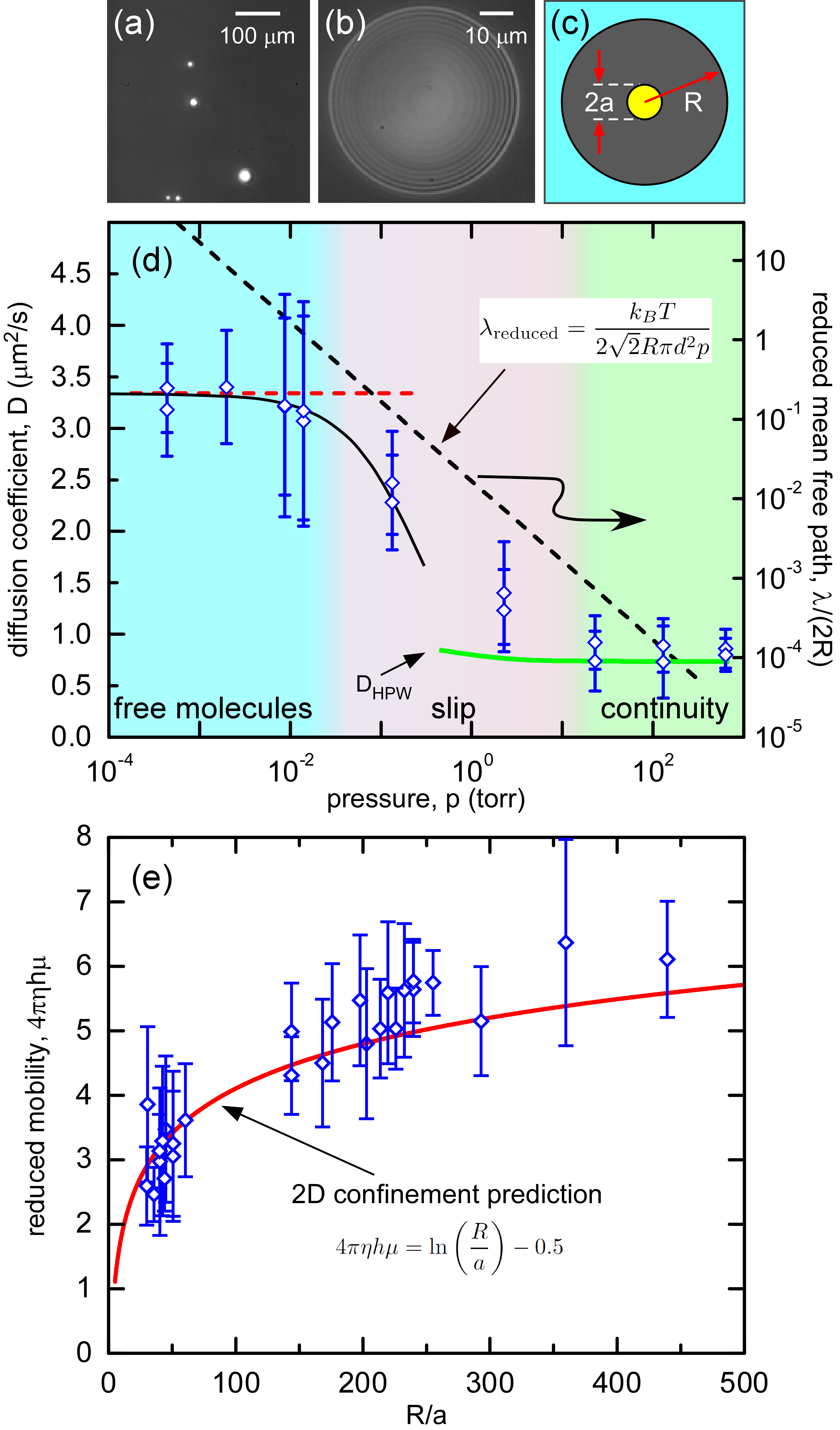}     
\caption{\label{fig2} Effect of surrounding air pressure on droplet diffusion. (a) Oil droplets on a three-layer, freely suspended liquid crystal film viewed in reflection. (b) Interference fringes in a large oil droplet. (c) Cartoon of an oil droplet of radius $a$ near the center of a film of radius $R$. (d) Diffusion coefficient of a single droplet ($a=8 \, \mu$m) near the center of a film ($R=2 \,$mm, $N=3$ layers) as a function of surrounding air pressure (symbols). The green curve corresponds to SD/HPW theory. The black curve shows the diffusion predicted by a model assuming free air molecules impinging on the film. The horizontal dashed red line shows the $2$D confinement limit predicted by Saffman for vanishingly small air viscosity and no-slip boundaries. The dashed black line shows the mean free path of the air molecules scaled by the film diameter ($\lambda_\mathrm{reduced}=\lambda/(2R)$) as a function of pressure. The  background shading indicates three distinct behavioral regimes corresponding to different air pressure ranges: free molecules, slip and continuity. (e) Reduced mobility of single oil droplets in a film in high vacuum as a function of reduced film radius. The model curve is Saffman's prediction for a $2$D fluid with no-slip boundaries.}
\end{figure}

The diffusion coefficient of a typical droplet near the center of the film is plotted as a function of ambient pressure in Fig.~\ref{fig2}(d). The green curve shows the corresponding SD/HPW theory with the air viscosity corrected for pressure \cite{Johnston1951}. The observed variation of diffusion coefficient is well described by this model at pressures close to atmospheric ($p\gtrsim70$ torr) but the experimental data deviate significantly from the theory at lower pressure, increasing monotonically as the pressure is reduced before saturating at very high vacuum, at the limit corresponding to $2$D boundary confinement (horizontal red dashed line in Fig.~\ref{fig2}(d)).
The observed behavior falls in three distinct hydrodynamic regimes: (1) Near atmospheric pressure ($p\gtrsim 70$~torr), the mean free path of the air molecules ($\lambda \sim 7 \,\mu$m) is much less than the diameter of the film. In this regime, the air may be regarded as a continuous fluid and SD/HPW theory gives diffusion $D$ with no adjustment parameters, using known air viscosity $\eta^{\prime}$, film viscosity $\eta$, and measured film thickness $h$, and the measured hydrodynamic radius (see Supporting Information) of the inclusions~\cite{Nguyen2010}. (2) Below about $70$~torr, the viscosity of the air decreases as the pressure falls, a phenomenon attributable to slip of the air layers \cite{Johnston1951} over the surfaces of the film and oil droplet. (3) At very low pressure ($p \lesssim 0.02$~torr), the mean free path is several mm long, a distance comparable to the diameter of the film. In this regime, the ambient air can be regarded as an ensemble of collisionless molecules that obey a Maxwell-Boltzmann velocity distribution \cite{Clark19751,Clark19752}.

In order to model the behavior of droplets at the lowest pressures, we may approximate the total drag force $F$ as the sum of two terms, one arising from confinement by the boundaries and the other due to friction from the air, or $F=F_b + F_\mathrm{air}$. The confinement term is given by $F_b = {4\pi\eta hU}/{(\ln({R}/{a}) - 0.5)}$ \cite{Saffman1975}. The air drag on an inclusion moving in the film at speed $U$ depends on both direct frictional force resulting from the impingement of air molecules on the inclusion, and on indirect frictional forces resulting from changes of streamlines in the film caused by collisions with air molecules. Calculations based on kinetic theory \cite{Epstein1924} indicate that the unit frictional force as a function of droplet speed $U$ and surrounding air pressure $p$ may be written $F_\mathrm{air} = p\sqrt{\pi m/(2kT)} \, U$, where $m$ is the mass of an air molecule, $k$ the Boltzmann constant, and $T$ the temperature. The net inverse droplet mobility may be written as $1/\mu = 1/\mu_{b} + 1/\mu_\mathrm{air}$. Since $\mu_b$ is independent of pressure, the mobility can be expressed in the form $\mu = 1/(\mu_b^{-1}+ \mathrm{const}\times p)$, where the constant can be found by fitting the experimental data at low pressure. This model is plotted as the black curve in Fig.~\ref{fig2}(d).

In an ideal $2$D fluid of finite size, therefore, the only drag experienced by a disk-like inclusion should come from confinement forces arising from long-range hydrodynamic interactions with the fluid boundaries. Our experiments confirm that in high vacuum ($p \lesssim 0.003$~torr), the frictional drag from the remaining air molecules is much smaller than the hydrodynamic confinement force and can be neglected. In this regime, the freely suspended SmA liquid crystal film approaches a true $2$D fluid and exhibits purely $2$D hydrodynamics. To verify that we were really in the $2$D limit, we analyzed the Brownian motion of droplets of different sizes in films of different radii under high vacuum. The reduced mobility $m = 4\pi\eta h \mu$ of these inclusions as a function of reduced film radius $R/a$ is plotted in Fig.~\ref{fig2}(e). The observed mobility follows the predictions of SD theory quite well, increasing logarithmically with inclusion size as expected for a system with $2$D hydrodynamic behavior. The observed mobilities are slightly larger than predicted by the model, an effect which might be due to deviations from ideal, no-slip boundary conditions resulting from the presence of a meniscus~\cite{Picano2000}.

\begin{figure}[tbh]
\centering
\includegraphics[width=7.5cm]{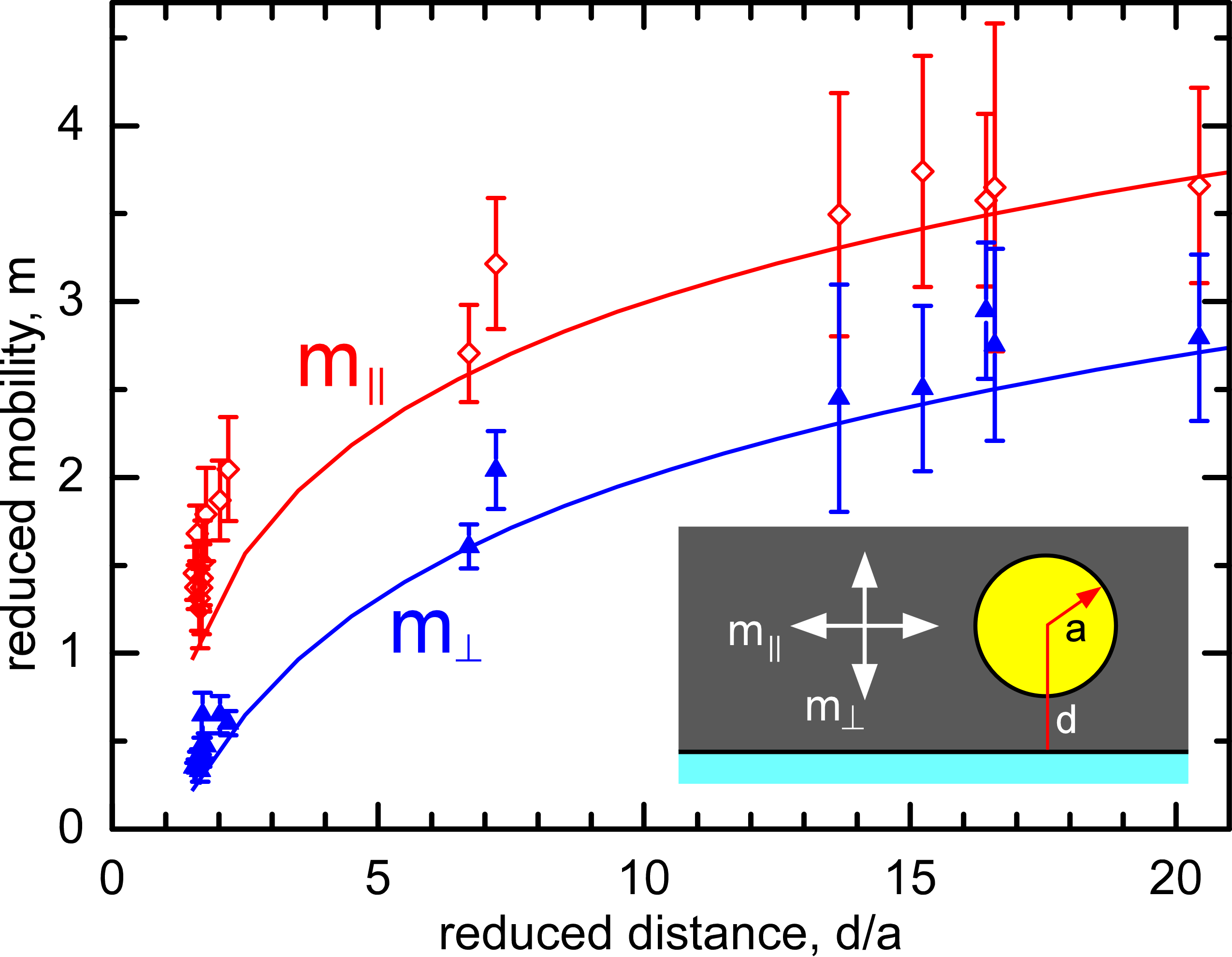}
\caption{\label{fig3} Reduced mobility of oildrop inclusions diffusing parallel (red) and perpendicular (blue) to a straight film boundary in high vacuum.  $a$ is the radius of the inclusion and $d$ the distance from the boundary. The model curves are analytical predictions from Eq.~\ref{eq:jeffrey}. The two mobilities are different, in agreement with theory, increasing logarithmically with distance from the boundary as expected for pure $2$D hydrodynamics.}
\end{figure}

In both $3$D~\cite{Banerjee2005, Lele2011} and $2$D fluids, inclusions near a rigid boundary experience a ``wall effect'' which reduces their mobility. To study the $2$D ``wall effect", in the $1940s$, White measured the drag on metal wires falling sideways through viscous liquids confined between two vertical bounding walls and found that at low Reynolds number, the presence of the walls affected the mobility of the wires even when they were many hundreds of wire diameters away, with the mobility depending logarithmically on the ratio of wall separation to  wire radius \cite{White1946}. Recent measurements of inclusion mobility in very thick smectic films, in which the Saffman length is greater than the film size and the influence of the air is relatively small, also showed the effects of the boundary \cite{Eremin2011}.

Eliminating the environmental drag on a thin smectic film by removing the ambient air seemed a promising way of studying wall effects in a true $2$D fluid.
We therefore measured the mobilities of included oil droplets both parallel and perpendicular to a straight boundary in high vacuum. The experimental observations, shown in Fig.~\ref{fig3}, were compared with the model of Jeffrey and Onishi \cite{Jeffrey1981}, who solved the Navier-Stokes equations for $2$D flow around a translating cylinder near a plane wall, assuming small Reynolds number flow. For translation respectively parallel and perpendicular to the wall, the predicted reduced mobilities are:
%
%
%
%
\begin{subequations}
    \label{eq:jeffrey}
\begin{eqnarray}
    m_\parallel & =  4\pi\eta h \mu_\parallel & =  \ln \left [\frac{d+\sqrt{d^{2}-a^{2}}}{a} \right ]
    \, , 	 \label{eq:parallel} \\
    m_\perp & = 4\pi\eta h \mu_\perp     & =   \ln \left [\frac{d+\sqrt{d^{2}-a^{2}}}{a} \right ]
     -\frac{\sqrt{d^{2}-a^{2}}}{d}  .
    \label{eq:perp}
\end{eqnarray}
\end{subequations}


\noindent
For large values of $d/a$, these expressions simplify to  $m_\parallel \approx \ln[2d/a]$  and $m_\perp \approx \ln[2d/a]-1$. In contrast to $3$D fluids, where the wall effect on mobility decays within a few dozen inclusion radii \cite{Carbajal2007}, the influence of the boundary  extends a long distance into a $2$D fluid and the hydrodynamic behavior of inclusions is predicted to remain anisotropic at large distances from the wall. Our experimental results confirm this behavior, as seen in Fig.~\ref{fig3}. The measured mobilities are on average slightly higher than the theory but are generally in good agreement, except very close to the wall. This may be due to deviations from true no-slip boundary conditions~\cite{Keh2008} at the meniscus, as mentioned previously.

In summary,  we have described the Brownian motion of single inclusions in freely suspended smectic~A liquid crystal films as the pressure of the surrounding air is reduced from one atmosphere to a high vacuum. The inclusion mobility was characterized in three hydrodynamic regimes: near atmospheric pressure, where diffusion follows HPW theory, in partial vacuum (the slip regime), and in high vacuum, where we observe motion limited by $2$D confinement effects. The parallel and perpendicular mobilities of an inclusion in high vacuum near the edge of the film increase logarithmically with distance from the boundary as predicted by theory, with an anisotropic character that persists far into the film. The observations suggest that thin, freely suspended smectic films in high vacuum are a nearly ideal experimental realization of a two-dimensional fluid. This work opens the way for more general hydrodynamic studies in the 2DIIN limit, of such phenomena as driven microrheological flow, high Reynolds number turbulence, energy cascades, and jets.

This work was supported by NASA Grant~NNX-13AQ81G and NSF MRSEC Grants~DMR-0820579 and DMR-1420736.

\bibliographystyle{apsrev}
\bibliography{vacuumref} 

\end{document}